\documentclass[%
 aip,
 amsmath,amssymb,
preprint,%
]{revtex4-2}
\usepackage{graphicx}
\usepackage{dcolumn}
\usepackage{bm}
\usepackage[english]{babel}
\usepackage[utf8]{inputenc}
\usepackage[T1]{fontenc}
\usepackage{mathptmx}
\usepackage{etoolbox}
\makeatletter
\def\@email#1#2{%
 \endgroup
 \patchcmd{\titleblock@produce}
  {\frontmatter@RRAPformat}
  {\frontmatter@RRAPformat{\produce@RRAP{*#1\href{mailto:#2}{#2}}}\frontmatter@RRAPformat}
  {}{}
}%
\makeatother
\usepackage{siunitx}

\newcommand{\lno}{LiNbO\textsubscript{3} }

\begin{document}


\title{Electro-optical modulation of light polarization in a nonlocal lithium niobate metasurface}
\author{Agostino Di Francescantonio}

 \altaffiliation[Also at ]{EPFL, Insitute of Mechanical Engineering, Laboratory of Nanoscience for Energy Technologies, Lausanne, Switzerland.}
 \affiliation{Politecnico di Milano, Physics Department, Milano, Italy. }
\author{Alessandra Sabatti}%
\affiliation{ 
ETH Zürich, Department of Physics, Institute for Quantum Electronics, Zürich, Switzerland.
}%
\author{Eleni Prountzou}
\affiliation{ 
ETH Zürich, Department of Physics, Institute for Quantum Electronics, Zürich, Switzerland.
}%
\author{Maria Antonietta Vincenti}
\affiliation{Università di Brescia, Department of Information Engineering, Brescia, Italy.}
\author{Luca Carletti}
\affiliation{Università di Brescia, Department of Information Engineering, Brescia, Italy.}
\author{Attilio Zilli}
\affiliation{Politecnico di Milano, Physics Department, Milano, Italy. }

\author{Michele Celebrano}
\affiliation{Politecnico di Milano, Physics Department, Milano, Italy. }

\author{Rachel Grange}
\affiliation{ 
ETH Zürich, Department of Physics, Institute for Quantum Electronics, Zürich, Switzerland.
}%
\author{Marco Finazzi}
\affiliation{Politecnico di Milano, Physics Department, Milano, Italy. }


\begin{abstract}
We report the experimental realization of a \lno metasurface for electro-optic modulation of light polarization in the telecommunication C-band. 
High quality factor quasi-bound states in the continuum are employed to enhance the modulation of the amplitude and phase of an impinging beam by a driving electric field, leading to efficient polarization rotation and conversion. We quantified modulation effects under a CMOS-compatible bias at 1 MHz frequency, achieving variations of $5\%$ in the Stokes parameters and a variation of the polarization ellipse angles of about \ang{3} for the transmitted light. These results demonstrate that dynamic polarization and phase modulation can be attained in a compact platform, highlighting the potential of high-quality resonant \lno nonlocal metasurfaces for enhanced light–matter interaction in subwavelength electro-optic devices.
\end{abstract}

\maketitle



\section{Introduction}
Optical metasurfaces -- artificial arrangements of nanostructures of sub-wavelength thickness -- provide a versatile  platform  for tailoring  light--matter interaction and manipulate the wavefront of propagating optical beams \cite{schulz_roadmap_2024}.
In particular, metasurfaces enable precise control over the polarization state of light, which can be achieved by shaping the phase and amplitude of the electromagnetic field \cite{overvig_dielectric_2019} or by exploiting nonlinear interactions\cite{wang_all-optical_2024}.
A wide variety of metasurface-based polarization devices have been reported in literature \cite{li_metasurface_2024}, exhibiting  advanced functionalities ranging from polarization conversion\cite{arbabi_dielectric_2015, fagiani_dualmode_2024} and  lensing \cite{khorasaninejad_metalenses_2016}
to imaging\cite{rubin_matrix_2019}, nonlinear generation of circularly polarized light\cite{luan_all-optical_2025}, and Bell states engineering\cite{ma_polarization_2023}, to name a few.

The reconfiguration of metasurfaces properties by means of external stimuli\cite{abdelraouf_recent_2022} -- typically thermal\cite{zograf_all-dielectric_2021}, electrical\cite{jung_rise_2024} or optical\cite{maiuri_ultrafast_2024}  -- is receiving a rapidly-growing interest. 
In particular, achieving  active control of a metasurface polarization response represents  a crucial step toward the realization of devices capable of  on-demand  polarization control and modulation, which can find applications in optical communications and LiDAR technology.

Materials exhibiting a tunable birefringence are commonly employed in commercial phase and polarization modulators to dynamically control the phase of a linearly-polarized optical wave.
These devices mostly rely on phase transitions in liquid crystals, photoelastic effects or the linear electro-optic (EO), namely Pockels' effect, which are typically driven by an electric signal.
The applied bias yields a change in the material refractive index, $\Delta n$, which translates into a dephasing $\Delta\delta=2\pi\,\Delta n \, l/\lambda$ introduced over a propagation length $l$ inside the medium.
Among the mentioned mechanisms for electrical modulation, the EO effect arising from a nonlinear interaction between a low-frequency electric field $\textbf{E}_\textsc{eo}$ and an optical field, produces a change in the \textit{i}-th component of the refractive index tensor $\Delta n_i = -\frac{1}{2}n_i^3r_{ij}E_{\textsc{eo},j}$, where $r_{ij}$ is an element of the material EO tensor.
The main advantages of the EO effect reside in the high modulation frequencies that can be achieved, potentially extending above hundreds of GHz \cite{wang_integrated_2018}, in the ease of integration in existing electro-optic   technologies and in the absence of Joule dissipation, making it an ideal candidate for high-speed and energy-efficient on-chip operations.
However, the limited refractive-index change $\Delta n$ achievable in routinely employed EO materials -- where the largest value of the coefficients $r_{ij}$ are typically smaller than \SI{100}{\pico\meter\per \volt} in currently available crystalline thin films -- imposes a long propagation length $l\gg\lambda$ (often in the millimeter range) and strong applied bias to obtain a sizable phase shift $\Delta\delta$ in EO devices. 
Optical metasurfaces offer a route to overcome such a constraint by exploiting the large field confinements and enhancements associated with photonic resonances.
Due to light trapping in the nanostructure, the effective interaction length is increased by the resonance quality-factor $Q$, boosting the values of $\Delta\delta$ even in sub-wavelength thickness samples. \\
Yet, the realization of even moderate \textit{Q} values ($10^3$ to $10^4$) has been often elusive in materials with large EO coefficients, such as lithium niobate (\lno), due to fabrication challenges. 
Recently, significant advancements in nano-fabrication techniques enabled the realization of optical devices characterized by a remarkable quality, allowing  the realization of nonlocal resonant modes, like guided-mode resonances and quasi-bound-states in the continuum (quasi-BICs), with increasing $Q$ values\cite{koshelev_nonlinear_2019}.
In this frame, EO devices based on \lno \cite{weiss_tunable_2022, damgaard-carstensen_nonlocal_2023}, and organic polymers \cite{zhang_high-speed_2023}, have demonstrated a steady performance improvement  over the last few years, culminating in intensity modulation efficiencies exceeding $\SI{1e-2}{\per\volt}$ and modulation frequencies in the GHz range\cite{damgaard-carstensen_highly_2025, di_francescantonio_efficient_2025, dagli_ghzspeed_2025, benea-chelmus_gigahertz_2022, soma_subvolt_2025}, which represent important milestones in the development of EO metasurfaces. 
Despite such progress, a dynamic control of light properties other than intensity remains unexplored in ultrathin EO modulators\cite{ding_electrically_2024}.
The experimental demonstration of polarization modulators has been limited to the microscale in the \si{\tera\hertz} regime \cite{zhang_tunable_2017} and, at the nanoscale, to other reconfiguration mechanisms, like liquid crystals \cite{yu_electrically_2021}, charge modulation\cite{wu_nearinfrared_2021} and thermo-optic effect \cite{bosch_polarization_2019}, while the use of the EO effect to control light polarization in optical metasurfaces has only been suggested by few theoretical studies \cite{wang_tunable_2023,hou_electrically_2024}.

In this letter we bridge this gap presenting a \lno EO metasurface that modulates the polarization of a transmitted beam in the telecommunication C-band. 
The device is based on a design that, by leveraging resonances with $Q>\SI{7e3}{}$, already demonstrated an intensity modulation efficiency of \SI{1.5e-2}{\per\volt} over a band of about \SI{1}{\giga\hertz} and a second-harmonic generation modulation of \SI{1.2e-1}{\per\volt}, see Ref.~\onlinecite{di_francescantonio_efficient_2025}. 
Here we show that the same architecture can exploit polarization-sensitive nonlocal resonances to modulate the phase difference of one specific linear polarization with respect to the perpendicular one.
We provide the first experimental demonstration of a fast EO metasurface that dynamically tunes the rotation angle and ellipticity of an incoming polarization by means of an applied radiofrequency field.
These results represent an important step toward the exploitation of EO effect in metasurfaces for a multidimensional control of light properties, including intensity, phase and polarization\cite{thureja_toward_2022}.    

\section{Design of the device}

\begin{figure}[h!]
    \centering
    \includegraphics[width=\linewidth]{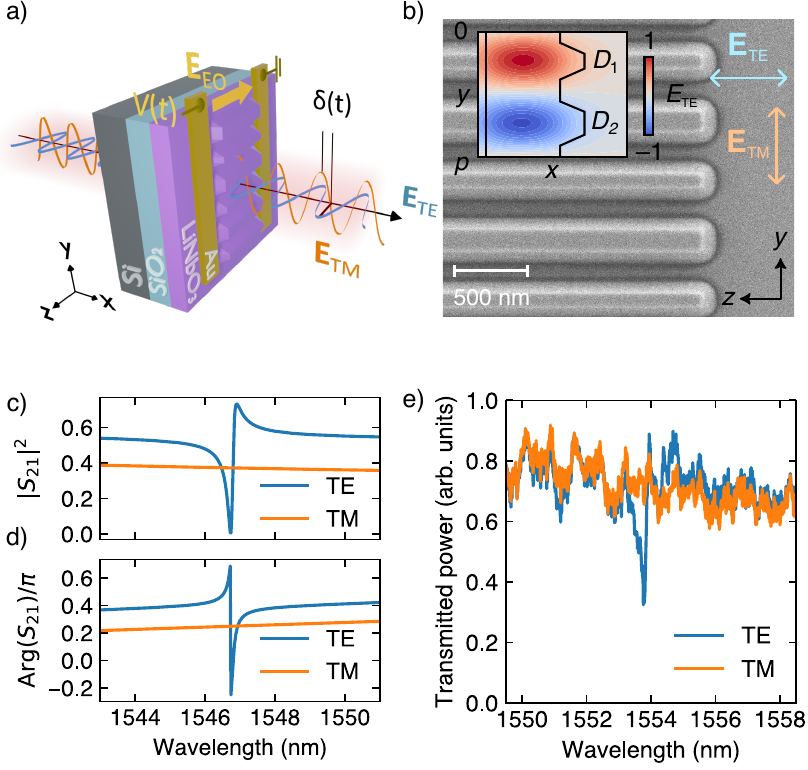}
    \caption{\textbf{a)} Illustration of a polarization modulator made by an $x$-cut lithium niobate (LiNbO\textsubscript{3}) thin film on insulator grating. 
     Any mixed polarization state (e.g., an input beam polarized at \ang{45} with respect to both the $x$ and $y$ axes) is dynamically tuned by an external bias $V(t)$, which generates an electric field $\mathbf{E}_\textsc{eo}$ and modifies the relative phase $\delta(t)=\delta_\textsc{te}(t)-\delta_\textsc{tm}$ between the transverse electric (TE) and magnetic (TM) components (blue and orange, respectively) of the optical field via the electro-optic effect. 
     \textbf{b)} scanning-electron micrograph of the investigated LiNbO\textsubscript{3} grating, with out-of-plane crystallographic $x$-axis, periodicity ${p = \SI{800}{\nm}}$, and asymmetry ${(D_1-D_2)/(D_1+D_2) = 0.2}$. The blue and orange arrows identify the TE and TM electric field components of the excitations.
     In the inset, COMSOL finite-elements simulation of the TE-polarized optical field distribution of the quasi-BIC mode.
     \textbf{c)},\textbf{d)} COMSOL finite-element simulation of \textbf{c)} the transmittance $|S_{21}| ^2$, where ${S}_{21}$ is the complex transmission coefficient, and  \textbf{d)} of the phase of $S_{21}$ for the two orthogonal excitations.
     \textbf{e)} Measured transmitted power through the metasurface for an impinging beam at normal incidence with the electric field polarized either TE or TM polarization. }
    \label{fig:SEM_transmission}
\end{figure}

Fig.~\ref{fig:SEM_transmission}a provides a schematic of a \lno metasurface modulating the light polarization by means of the EO effect
The realized device (see Fig.~\ref{fig:SEM_transmission}b), whose fabrication details can be found in Ref. \onlinecite{di_francescantonio_efficient_2025}, is realized from a \SI{500}{\nm}-thick $x$-cut \lno film on top of a \SI{2}{\micro\meter}-thick insulating SiO\textsubscript{2} layer.
The \lno film is patterned with asymmetric nanowires, aligned parallel to the \lno crystallographic \textit{z} axis, to form a one-dimensional periodic grating with in-plane broken symmetry.
The optical field impinging on the device can be conveniently decomposed as $\mathbf{E}={E}_\textsc{te}\,e^{i\delta_\textsc{te}}\hat{\mathbf{u}}_\textsc{te}+{E}_\textsc{tm}e^{i\delta_\textsc{tm}}\hat{\mathbf{u}}_\textsc{tm}$, where ${E}_\textsc{te,tm}$ are the transverse-electric (TE) and transverse-magnetic (TM) amplitudes, parallel and perpendicular to the nanowires (i.e., aligned to the \textit{z} and \textit{y} axis, see also Fig.~\ref{fig:SEM_transmission}b), respectively, and $\delta_\textsc{te,tm}$ are the corresponding phases.
Therefore, the light polarization is defined by the amplitude and phase relation between these two orthogonal components. 
The polarization of an arbitrary superposition of TE and TM components -- e.g., a linear polarization oriented at \ang{45} with respect to the \textit{z} axis -- can be modified by changing both their relative amplitude and relative phase (i.e., the retardance) $\delta=\delta_\textsc{te}-\delta_\textsc{tm}$.
In the presented metasurface, this is achieved by exploiting the in-plane birefringence of the \textit{x}-cut \lno, which produces a spectral mismatch between TE and TM narrowband resonances.
In other words, in a given wavelength interval, only one component of the optical field interacts with the metasurface, undergoing a wavelength-dependent amplitude variation and phase shift $\Delta \delta(\lambda)$.
The other one, instead, remains unaffected when crossing the sample. 
The polarization of the transmitted beam can thus be tuned by shifting the central wavelength $\lambda_0$ of the resonance through the change in the material refractive index induced by an electrical bias.

The metasurface has been designed to support a resonance excited by a TE optical field in the telecommunication C-band (i.e., \si{1530} to \SI{1565}{\nm}).
The biasing field is applied to the metasurface by in-plane gold electrodes positioned \SI{15}{\micro\meter} apart, perpendicularly to the nanowires.
According to numerical simulations\cite{di_francescantonio_efficient_2025}, it is approximately uniform within the \lno substrate and aligned parallel to the crystallographic \textit{z} axis. 
In this geometry, the variation of the refractive index for the TE optical field component is determined by the $r_{33}$ component of the \lno EO tensor, which is the largest one (\SI{35}{\pico\meter\per\volt}, see Ref.~ \onlinecite{holmes_evaluation_1983}). 
The applied field $\mathbf{E}_\textsc{eo}=E_\textsc{eo}\hat{\mathbf{u}}_z$ modifies the \lno extraordinary index $n_\mathrm{e}$ according to the expression\cite{fedotova_lithium_2022}  
\begin{equation}\label{eq:delta_e}
    \Delta n_\mathrm{e} = -\frac{1}{2}n_\mathrm{e}^3r_{33}E_\textsc{eo}.
\end{equation}
The narrowband character of the resonance increases the sensitivity to small $\Delta n_\mathrm{e}$ thanks to the not negligible spectral shift of the resonance central wavelength $\lambda_0$ with respect to the resonance linewidth $\delta\lambda$, that can be obtained even at moderate bias voltages.

The realized device supports high-\textit{Q} guided mode resonances arising from the original guided modes of the \lno slab upon translational symmetry reduction introduced by the patterned periodic nanowires\cite{huang_ultrahigh-q_2023, sun_infinite-q_2023}. 
In the case of a sub-wavelength grating characterized by a \textit{xz} mirror symmetry, a fundamental, symmetry-protected TE bound state in the continuum (BIC) with antisymmetric field pattern (see the inset in Fig.~\ref{fig:SEM_transmission}b) and theoretically null radiative losses is present for a vanishing in-plane wavevector $k_{\parallel}=0$.
As in Ref.~\onlinecite{di_francescantonio_efficient_2025}, to allow radiation coupling to this mode at normal incidence, the \textit{xz} mirror symmetry is perturbed through a slight asymmetry in the nanowires width, turning the BIC into a quasi-BIC, with a finite quality-factor.
This is achieved by realizing a grating with a subwavelength period $p=\SI{800}{\nm}$, characterized by an asymmetry factor $\alpha={(D_1-D_2)/(D_1+D_2)}=0.2$, with $D_1$ and $D_2$ equal to the width of the reliefs in each unit cell (see Fig.~\ref{fig:SEM_transmission}b), and filling factor $\mathrm{FF} = {1- (D_1+D_2)/p}=0.3$.
According to numerical simulations, the resulting quasi-BIC, excited by TE polarization, emerges at the telecommunication wavelength of $\lambda_0 \simeq \SI{1547}{\nm}$ (see Fig.~\ref{fig:SEM_transmission}c) with a theoretical quality-factor $Q=10^4$.
It is worth mentioning that, despite often realized in integrated photonic devices -- including microring resonators, microcavities, and two-dimensional photonic crystals \cite{zhang_monolithic_2017, akahane_high-q_2003, sanvitto_observation_2005} --, aiming for a larger \textit{Q} can be impractical in metasurfaces, as it results in increasingly inefficient coupling to propagating fields.
As evidenced by the numerical simulations in Fig.~\ref{fig:SEM_transmission}c, the quasi-BIC gives rise to a sharp Fano resonance in the sample transmittance $|S_{21}|^2$, together with a steep variation of the phase of the of the transmission coefficient $S_{21}$ (see Fig.~\ref{fig:SEM_transmission}d).
Conversely, the transmission spectrum of the TM polarization appears featurless in the same spectral range.

The sample resonant properties are characterized in a transmission spectroscopy setup with detection unsensitive to the light polarization.
The sample is illuminated with a continuous-wave tuneable diode laser in the communication C-band and the transmitted power is collected by a photodiode (see Supplementary Information, S1).
Fig.~\ref{fig:SEM_transmission}e shows the transmission spectra obtained as a function of the wavelength of a TE- or TM-polarized beam.
The quasi-BIC mode is observed as a pronounced dip in the TE transmission spectrum with an asymmetric profile.
The resonance central wavelength $\lambda_0\simeq\SI{1553.80}{\nm}$ and linewidth $\delta\lambda = \SI{0.17}{\nm} $ are retrieved upon fitting the mode profile with a Fano lineshape\cite{limonov_fano_2021}, corresponding to a quality factor $Q=\lambda_0/\delta\lambda\simeq9000$.
These values are in good agreement with the numerical prediction shown in Fig~\ref{fig:SEM_transmission}c, suggesting an excellent fabrication quality.
On the other hand, the TM optical component doesn't show any resonant feature in the same spectral region.
We mention that the raw transmission spectra were affected by strong Fabry--Pérot oscillations compatible with a thickness of the Si substrate of about \SI{400}{\nm}. 
Therefore, we fitted the original data with the transmissivity of a Fabry--Pérot cavity, removing the corresponding oscillating contribution to obtain the spectra reported in Fig.~\ref{fig:SEM_transmission}e (see Section S2, Supplementary Information).

\begin{figure*}
    \centering
    \includegraphics[width=\linewidth]{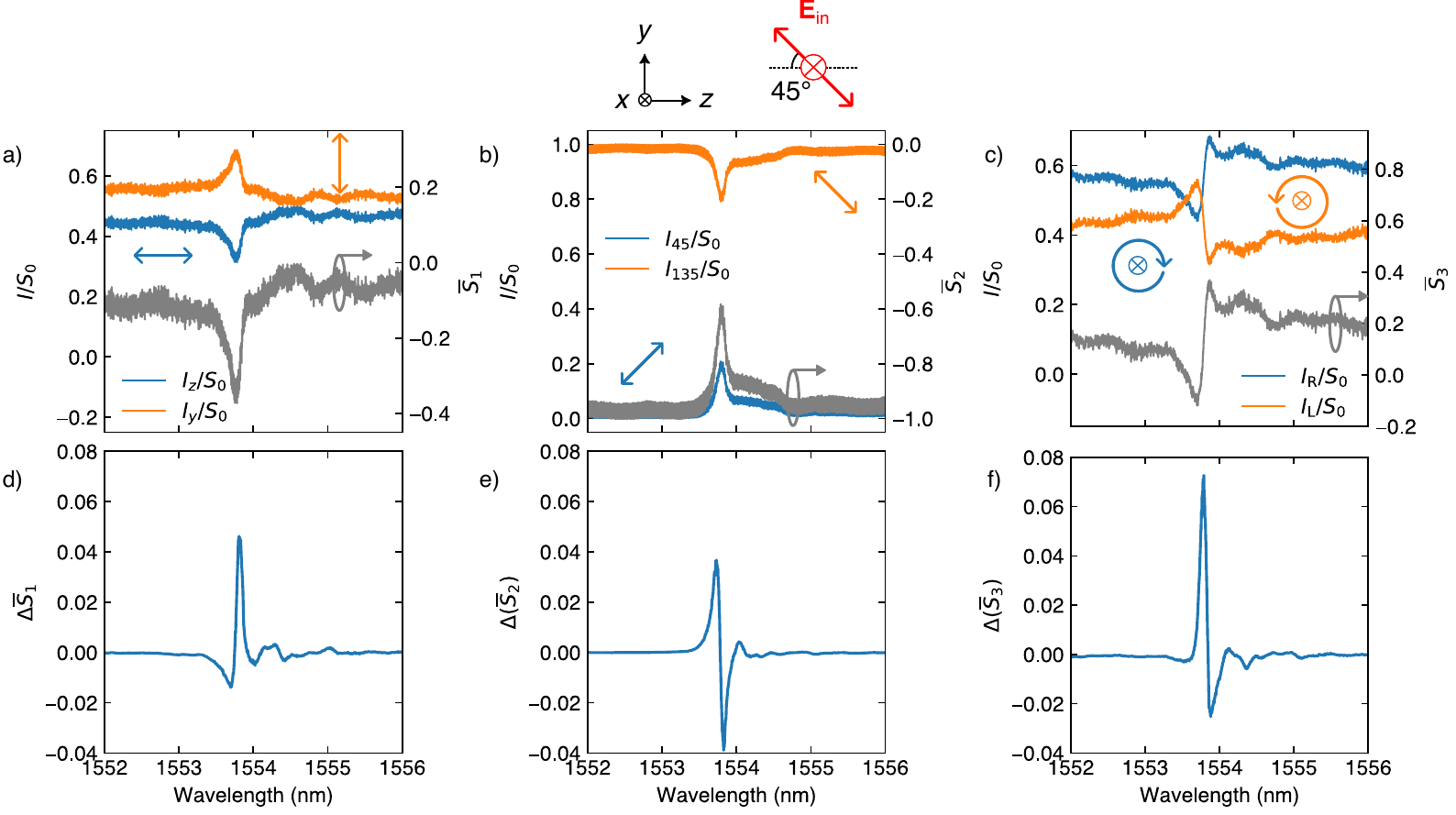}
    \caption{Across-resonance static and dynamic characterization of the transmitted polarization for an impinging  beam linearly polarized at \ang{135} with respect to the LiNbO\textsubscript{3} \textit{z} axis (see sketch on top). 
    \textbf{a)}--\textbf{c)} Plot of the orthogonal polarization states in blue (orange), identified by arrows in each panel, in the linear \textit{z} (\textit{y}), linear diagonal \ang{45}(\ang{135}) and circular left (right) basis, respectively.
    The intensities in each panel are normalized to the total intensity $S_0=I_a+I_b$, where $I_{a,b}$ are the two measured projections in the corresponding basis. 
    In grey, the corresponding normalized Stokes parameters $\overline{S}_1$, $\overline{S}_2$ and $\overline{S}_3$.  \textbf{d)} -- \textbf{f)} Variation of the normalized Stokes parameters upon application of a sinusoidal driving voltage with peak-to-peak amplitude $V_\mathrm{pp} = \SI{10}{\volt}$ and frequency $f_\mathrm{mod} = \SI{1}{\mega\hertz}$.}
    \label{fig:45deg_stokes}
\end{figure*}
\section{Results and discussion}
\subsection{Polarization characterization and modulation}
The light polarization state is conveniently expressed in terms of the Stokes parameters $S_1$, $S_2$ and $S_3$, representing the degree of polarization projected onto three orthogonal polarization bases, that can be can be determined from light observables\cite{hecht_optics_2017}. 
In particular, $S_1 = \langle Z\rangle-\langle Y\rangle$, $S_2 = \langle D\rangle-\langle A\rangle$ and $S_3 = \langle R\rangle-\langle L\rangle$, where $\langle\cdot\rangle$ are the time-averaged intensities  projected on the \textit{z} or \textit{y} axes, on the direction at $\pm\ang{45}$ with respect to these axes, and on the circular left or right polarization states, respectively.
The total intensity is  defined as $S_0=\langle a\rangle+\langle b\rangle$, where $\langle a,b\rangle$ are the two orthogonal projections in any given basis. 
The rotation and retardance variation effects imparted by the sample on the light polarization are assessed by evaluating the Stokes parameters of the transmitted radiation for different input polarization states.
The transmitted $S_1$ and $S_2$ components are determined by means of a polarization analyzer placed before the detector, while for  measurements in the circular basis $S_3$ a quarter-wavelength retarder is introduced before the analyzer.
As will be discussed in the following section, to completely characterize the device properties with intensity measurements, at least one input polarization in each basis set is required \cite{reddy_measuring_2014}.
As an example, the case of an antidiagonal input polarization state $\langle A\rangle$, with an associated Stokes parameters $S_1 = 0$, $S_2=-1$ and $S_3=0$, corresponding to a linearly polarized impinging beam at an angle of \ang{135} with respect to the \lno \textit{z} axis, is presented below. 
The characterization of other polarizations can be found in Section S1 of the Supplementary Information.
Figures~\ref{fig:45deg_stokes}a--c) show the normalized transmitted intensities projected onto the orthogonal axes for each of the three polarization bases and the corresponding normalized Stokes parameters $\overline{S}_i=S_i/S_0$.
As discussed above, the dispersive birefringence and transmittance of the metasurface, when probed at the TE resonant mode, induces  variations of the beam phase and amplitude, respectively.
As a consequence, the Stokes parameters exhibit a strong modulation across the resonance. 
In particular, the attenuation of the TE component produces a rotation of the polarization axis toward the \textit{y} axis, as shown by the dip in the $\overline{S}_1$ parameter (see Fig.~\ref{fig:45deg_stokes}a) and by the corresponding increase of the diagonally-polarized intensity $I_{45}$ (see Fig.~\ref{fig:45deg_stokes}b). 
Moreover, the variable phase shift $\delta_\textsc{te}(\lambda)$ results in a change in the $\overline{S}_3$ parameter, thereby modifying the polarization ellipticity (see Fig.~\ref{fig:45deg_stokes}c). 

We perform a dynamic modulation by applying a sinusoidal voltage $V(t)=\frac{V_\mathrm{pp}}{2}\sin{(2\pi f_\mathrm{mod}t)}$.
Under the assumption of a small EO-induced resonant shift, the transmitted power detected by the photodiode can be expressed  as $P(t) = P_\textsc{dc}+P(f_\mathrm{mod})\sin{(2\pi f_\mathrm{mod}t)}$. 
The amplitude of the  oscillating component, $P(f_\mathrm{mod})$, is experimentally determined via lock-in detection, following the procedure described in ref.~\onlinecite{di_francescantonio_efficient_2025}.
The device exhibits a tuning sensitivity of the resonance central wavelength of \SI{5.6}{\nm\per\volt}, allowing an amplitude modulation efficiency of \SI{1.5e-2}{\per\volt} in the small bias regime\cite{di_francescantonio_efficient_2025} -- i.e., for applied $V_\mathrm{pp}<\SI{10}{\volt}$.
To estimate the EO-induced polarization change, we first evaluate the shifted spectrum $P_i(\lambda,V=V_\mathrm{pp}/2)=P_{i,\textsc{dc}}(\lambda)+P_i(\lambda,f_\mathrm{mod})$ for $i$ running over the six projections displayed in Fig.~\ref{fig:45deg_stokes} a--c.
Then, we compute the corresponding normalized Stokes parameters
\begin{equation}\label{eq:Stokes_variation}
    \overline{S}_i(\lambda,V_\mathrm{pp}/2)= \frac{P_a(\lambda,V_\mathrm{pp}/2)-P_b(\lambda,V_\mathrm{pp}/2)}{S_0(\lambda,V_\mathrm{pp}/2)} ,
\end{equation}
where $P_{a,b}$ are the spectra of the two orthogonal polarizations in the considered basis.
The retrieved Stokes parameter modulation amplitudes $\Delta \overline{S}_i(\lambda)=\overline{S}_i(\lambda,V_\mathrm{pp}/2)-\overline{S}_i(\lambda,0)$ for $V_\mathrm{pp}= \SI{10}{\volt}$ at a modulation frequency of $f_\mathrm{mod}=\SI{1}{\mega\hertz}$ are shown in Figs.~\ref{fig:45deg_stokes}d--f.

\begin{figure}
    \centering
    \includegraphics[width=0.9\linewidth]{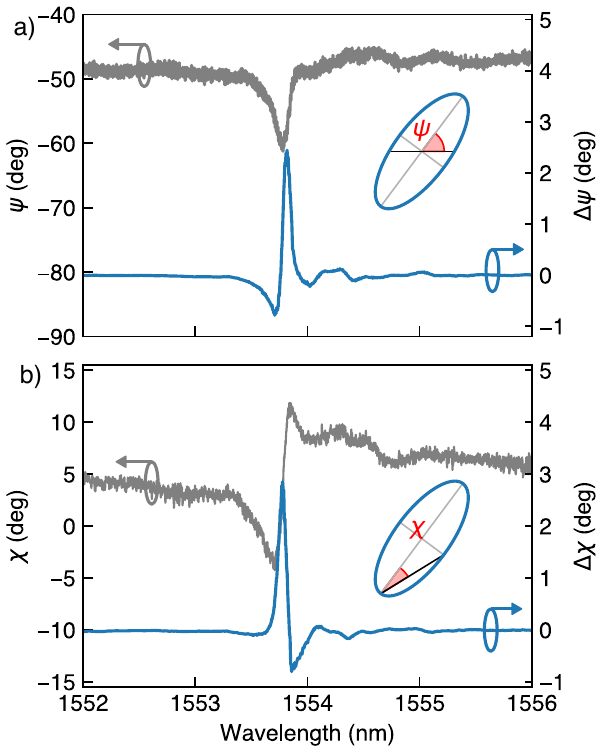}
    \caption{Static dispersion (in grey) and dynamic modulation (in blue) of the \textbf{a)} ellipse rotation angle $\psi$ and \textbf{b)} ellipticity angle $\chi$ of a \ang{135}-polarized beam, depending on the laser wavelength. 
    The sinusoidal modulating voltage is characterized by $V_\mathrm{pp}= \SI{10}{\volt}$ and $f_\mathrm{mod}=\SI{1}{\mega\hertz}$.
    }
    \label{fig:polarization ellipse}
\end{figure}
\begin{figure}
    \centering
    \includegraphics[width = .8\linewidth]{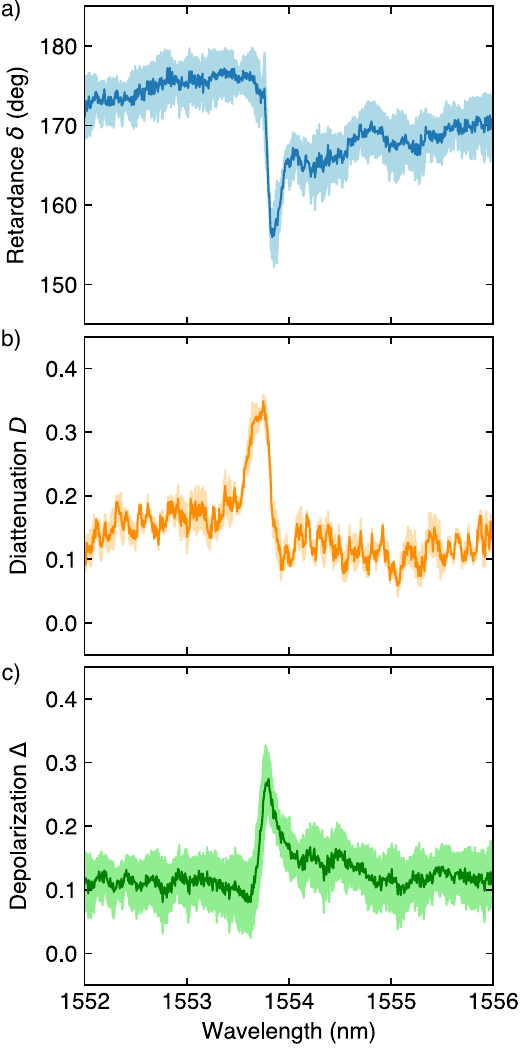}
    \caption{Spectra of \textbf{a} the retardance $\delta$,  \textbf{b} the diattenuation magnitude  $D$ and \textbf{c} the depolarization magnitude $\Delta$. The solid lines correspond to the average values obtained by calculating  the  Mueller matrices from four different polarization combinations, while the shaded areas are the corresponding standard deviations. }
    \label{fig:r_D_Delta}
\end{figure}
To better visualize  the effect of the resonance and of the EO-induced modulation on the light polarization, we estimated the quantities defining the polarization ellipse, which are the rotation angle $\psi$ and the ellipticity angle $\chi$ (see the illustrations inside Fig.~\ref{fig:polarization ellipse}a and b, respectively), which depend on the Stokes parameters according to the following expressions:
\begin{equation}
    2\psi = \arctan{\frac{\overline{S}_2}{\overline{S}_1}}, \quad 2\chi = \arctan{\frac{\overline{S}_3}{\sqrt{\overline{S}_1^2+\overline{S}_2^2}}},
\end{equation}
while the handedness (or helicity) of the light is determined by the sign of $\chi$, with a positive sign standing for right handed circularly polarized light and a negative sign for left handed circularly polarized light.
The dependence of $\psi$ and $\chi$ on wavelength is plotted as a grey line in Fig.~\ref{fig:polarization ellipse}.
The maximum rotation imparted by the sample is \ang{15}, caused by the attenuation of the \textit{z} component of the optical field. 
A change in ellipticity of \ang{15} is also observed, with a reversal of the helicity when crossing $\chi=0$ value (fully-linear polarization).
Concerning the modulation of the polarization ellipse, an absolute change of about \ang{3} is obtained for both the angles for $V_\mathrm{pp}=\SI{10}{\volt}$ (see Fig.~\ref{fig:polarization ellipse} in blue) and wavelengths around \SI{1553.8}{\nm}, corresponding to a modulation sensitivity of $\ang{0.6}\si{\per\volt}$.
As a final remark, the frequency response of this sample was characterized in ref.~\onlinecite{di_francescantonio_efficient_2025}, resulting in a modulation bandwidth of \SI{800}{\mega\hertz}. 
The same maximum modulation speed is expected also in this case, since the device time constant was only limited by the electrodes capacitance.
\subsection{Evaluation of the metasurface Mueller matrix}
We investigated how the device modulates the light polarization by inducing amplitude and phase variations in the TE electric field when  the metasurface is excited at resonance (see Fig.~\ref{fig:SEM_transmission}).
We now discuss how the phase information can be retrieved from the polarization measurement presented above.
Note that the Stokes parameters are obtained by measuring light intensities, eventually losing the information about the phase.
Therefore, the characterization of the Stokes parameters with a single input polarization is not sufficient to retrieve this information.
To circumvent this problem, we determine the real-valued $4\times 4$ Mueller matrix $\mathbf{M}$, which relates the input and output Stokes vectors $\mathbf{S}=\begin{bmatrix}
S_0&S_1&S_2&S_3
\end{bmatrix}^\mathrm{T}$ and fully describes the effect of the device on the light polarization. 
Here, we note that the total intensity $S_0$ has been averaged over its definitions in the three orthogonal polarization bases.
$\mathbf{M}$ represents a linear map in $\mathbb{R}^4$ containing 16 independent elements; therefore it is univoquely determined by how it transforms a basis of $\mathbb{R}^4$.
Let us pick any four linearly independent input Stokes vectors -- for example $\mathbf{S}_\textsc{te}$, $\mathbf{S}_{45}$, $\mathbf{S}_{135}$ and $\mathbf{S}_\textsc{l}$.
The input matrix constructed by arranging these vectors as its columns
$\mathbf{\tilde{S}}_\mathrm{in}=\begin{bmatrix}
\mathbf{S}_\textsc{te}&\mathbf{S}_{45}&\mathbf{S}_{135}&\mathbf{S}_\textsc{l}
\end{bmatrix}_\mathrm{in}$ is therefore invertible. 
We point out that, as anticipated above, at least one vector in each of the three polarization bases is required to satisfy this condition \cite{reddy_measuring_2014}. 
We measured the corresponding output states, which can be similarly arranged in the output matrix $\tilde{\mathbf{S}}_\mathrm{out}$.
Eventually, the Mueller matrix of the samples is then determined as
\begin{equation}\label{eq:mueller_matrix_sample}
\mathbf{M}=\mathbf{\tilde{S}}_\mathrm{out}\mathbf{\tilde{S}}_\mathrm{in}^{-1}.
\end{equation}
We should recall that the metasurface is not simply dephasing $\mathbf{E}_\textsc{te}$ and $\mathbf{E}_\textsc{tm}$ but is also strongly attenuating the transmitted $\mathbf{E}_\textsc{te}$ when illuminated at resonance with the quasi-BIC.
Therefore, $\mathbf{M}$ does not describe a simple phase retarder as it would be for a flat transmission spectrum, and it should be decomposed to disentangle the changes in phase and amplitude of the TE component.
We exploit an extensively used approach for Mueller matrix decomposition, consisting in a algorithm developed by Lu and Chipman\cite{lu_interpretation_1996}, which allows writing an arbitrary Mueller matrix as 
\begin{equation}\label{eq:luChipman_decomposition}  \mathbf{M}=\mathbf{M}_\Delta\mathbf{M}_\mathrm{r}\mathbf{M}_\mathrm{d},
\end{equation}
where the three matrices $\mathbf{M}_\Delta$, $\mathbf{M}_r$ and $\mathbf{M}_d$ account for depolarization (i.e., polarization scrambling), retardance and diattenuation (i.e., change in the relative phase and amplitude between two orthogonal polarizations) properties, respectively.  
Relevant details on the decomposition algorithm are discussed in Section S4 of the Supplementary Information.
Three meaningful quantities describing the action of the device on light polarization can be retrieved from the three matrices of Eq. \eqref{eq:luChipman_decomposition}:
\begin{subequations}
\label{eq:retardance1}
\begin{equation}
    \delta=\cos^{-1}{\biggl(\frac{\mathrm{Tr}\{\mathbf{M}_\mathrm{r}\}}{2}-1\biggr)},
\end{equation}
\begin{equation}
        D = \frac{1}{m_{00}}\sqrt{\sum_{j=1}^3{m_{0j}^2}},
\end{equation}
\begin{equation}
    \Delta = 1-\frac{1}{3}|\mathrm{Tr}\{\mathbf{M}_\Delta\}-1|,
\end{equation}

\end{subequations}
where $\delta=\delta_\textsc{te}-\delta_\textsc{tm}$ is the applied retardance, $D$ is the diattenuation magnitude, defined by the elements of the first row of \textbf{M},  $m_{0,j}$, and $\Delta$ is the depolarization magnitude. 
Thanks to the redundant set of polarization states that we acquired, the quantities defined in Eq.~\eqref{eq:retardance1} have been determined by averaging the decomposition results obtained from  four independently measured Mueller matrices.
Their dispersion across the resonance is plotted in Fig.~\ref{fig:r_D_Delta}.
The metasurface modifies the relative phase delay $\delta$ between the TE and TM components by about \ang{-23} across the resonance (see Fig.~\ref{fig:r_D_Delta}a), with a dispersion profile that qualitatively reproduces the simulated one (see Fig.~\ref{fig:SEM_transmission}d).
It is worth noting that, while the abrupt phase change is preserved, the absolute variation across the resonance is reduced with respect to Fig.~\ref{fig:SEM_transmission}d. 
This effect can be ascribed to the ion etching process\cite{gao_electro-optic_2021}, which deteriorates the material quality introducing optical losses \cite{geiss_photonic_2014}.
Fig.~\ref{fig:r_D_Delta}b depicts the diattenuation magnitude $D$, which is enhanced by the reduction of transmitted amplitude of $\mathbf{E}_\textsc{te}$ upon interaction with the metasurface resonance -- the element $m_{01}/m_{00}$ shows a sharp dip on resonance, see Fig.~S4b of the Supplementary Information.
Finally, Fig.~\ref{fig:r_D_Delta}c evaluates the depolarization, that is also increased on resonance. 
\begin{figure}
    \centering
    \includegraphics[width=0.9\linewidth]{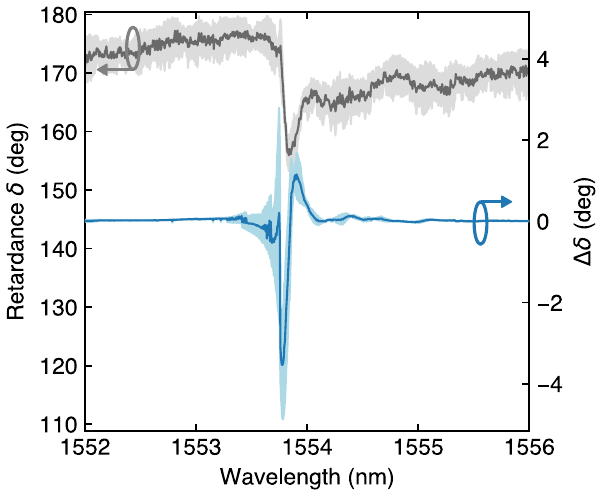}
    \caption{Wavelength dependence of the retardance $\delta =\delta_\textsc{te}-\delta_\textsc{tm}$ (in gray) and of its variation $\Delta \delta=\delta(V_\mathrm{pp}/2)-\delta(0)$ (in blue)  upon application of a sinusoidal driving voltage with peak-to-peak amplitude $V_{\mathrm{pp}}=\SI{10}{\volt}$ and frequency $f_\mathrm{mod} = \SI{1}{\mega\hertz}$. The difference $\Delta\delta$ has been calculated by evaluating $\delta$ from the metasurface Mueller matrices at \SI{0}{V} and $V_\mathrm{pp}/2$.
    The solid lines are obtained by averaging the results of the decomposition algorithm applied to four different polarization combinations. The shaded areas are the corresponding standard deviations.}
    \label{fig:retardance_variation}
\end{figure}
We ascribe this effect to an enhancement of the polarization scrambling introduced by scattering from defects and surface roughness, which are byproducts of the polishing treatment of sample back surface.

Eventually, we evaluate the capability of the device to modulate the relative phase between the TE and TM electric field components by determining the change in the retardance driven by the applied electric field.
Figure~\ref{fig:retardance_variation} shows the amplitude of the retardance modulation $\Delta\delta =\delta(\lambda, V_\mathrm{pp}/2)-\delta(\lambda,0)$  obtained by applying a sinusoidal bias with $V_\mathrm{pp}=\SI{10}{\volt}$. 
The absolute phase variation reaches approximately \ang{3}, which is in agreement with the measured change in ellipticity of Fig.~\ref{fig:polarization ellipse}b.
Notably, the same phase modulation would require a propagation length around $\SI{1}{\mm}$ assuming a refractive index modulation $\Delta n =10^{-4}$ to $10^{-5}$ in a non-resonant device.

\section{Conclusions}
In this work we presented the first experimental demonstration of a \lno metasurface capable of electro-optical modulation of light polarization.
A narrowband quasi-BIC resonance in the telecommunication C-band, selectively excited by a TE-polarized beam, induces a sizable attenuation of that component and a corresponding phase variation, resulting in a controllable in polarization rotation and conversion.
These effects were quantitatively characterized through full Stokes polarimetry of the transmitted light as a function of wavelength and incident polarization.
A dynamic polarization modulation was observed by applying a sinusoidal bias, resulting in an absolute variation of the Stokes parameters of the order of 0.05 for a peak-to-peak voltage amplitude $V_\mathrm{pp}=\SI{10}{\volt}$ at a modulation frequency of \SI{1}{\mega\hertz}.
These changes translated into a modulation of the polarization ellipse, with changes of both the rotation angle and ellipticity of the polarization ellipse up to \ang{3}, yielding to a sensitivity of \ang{0.6}\si{\per\volt}.
Furthermore, from the measured Stokes parameters the device Mueller matrix was retrieved, with an estimated phase variation of the TE-polarized electric field of \ang{3} for $V_\mathrm{pp}=\SI{10}{V}$. 
The achieved phase modulation extends the functionalities of nanoscale EO modulators from amplitude to phase and polarization control,  demonstrating the crucial role of high-\textit{Q} resonances in enhancing light-matter interaction in sub-wavelength optical devices.
This work extends the functionality of electro-optical modulators beyond amplitude control to simultaneous phase and polarization modulation, enabling compact, high-speed polarization control for optical communications, LiDAR, and emerging quantum and neuromorphic photonic platforms.

\bibliography{References1}
\end{document}